\documentclass[a4paper]{article}
\usepackage{xcolor}
\usepackage{soul}
\usepackage{INTERSPEECH2022}

\title{Device-Directed Speech Detection: Regularization via Distillation \\for Weakly-Supervised Models}
\name{Vineet Garg$^{1\ast}$, Ognjen (Oggi) Rudovic$^{1\ast}$,Pranay Dighe$^{1\ast}$,\\{\it Ahmed H. Abdelaziz$^{1}$, Erik Marchi$^{1}$, Saurabh Adya$^{1}$, Chandra Dhir$^{2\dagger}$, Ahmed Tewfik$^{1}$}
\thanks{$^{\ast}$Equal contribution.}
\thanks{$^{\dagger}$ Work done while at Apple.}}
\address{$^{1}$Apple, $^{2}$JPMorgan Chase \& Co.}
\email{vineetgarg@apple.com}

\begin{document}

\maketitle

\begin{abstract}
We address the problem of detecting speech directed to a device that does not contain a specific wake-word. Specifically, we focus on audio coming from a touch-based invocation. Mitigating virtual assistants (VAs) activation due to accidental button presses is critical for user experience. While the majority of approaches to false trigger mitigation (FTM) are designed to detect the presence of a target keyword, inferring user intent in absence of keyword is difficult. This also poses a challenge when creating the training/evaluation data for such systems due to inherent ambiguity in the user's data. To this end, we propose a novel FTM approach that uses {\it weakly-labeled} training data obtained with a newly introduced data sampling strategy. While this sampling strategy reduces data annotation efforts, the data labels are noisy as the data are not annotated manually. We use these data to train an acoustics-only model for the FTM task by regularizing its loss function via knowledge distillation from an ASR-based (LatticeRNN) model. This improves the model decisions, resulting in 66\% gain in accuracy, as measured by equal-error-rate (EER), over the base acoustics-only model. We also show that the ensemble of the LatticeRNN and acoustic-distilled models brings further accuracy improvement of 20\%.

\end{abstract}

\section{Introduction}
Virtual Assistants (VAs) are becoming a fundamental part of the interaction between users and smart devices, such as mobile phones, smart speakers, and watches, among others.
While VAs typically rely on detecting a wakeword or trigger phrase for invocation, another common method of interacting with VAs is to press a physical button on the smart-device, or tap on the touchscreen. This touch-based invocation method does not suffer from false alarms due to wakeword-like background speech, but it is prone to wrongly invoking the VA due to the button getting accidentally pressed (e.g. when the smartphone is in the user's pocket). In such cases, an unintended (the terms (un)intended and (un)directed are used interchangeably in this paper.) invocation of the VA may be recognized and affect the user experience~\cite{2020unacceptable}. In addition, such accidental invocation of the VA may undesirably interrupt the user and result in an overall unpleasant user experience. We refer to the task of device-directed speech detection in terms of detecting and rejecting such unintended touch-based invocations of the VA as False Trigger Mitigation (FTM). While various successful FTM approaches have been proposed in the literature for wake-word invocation~\cite{Kumar2020BuildingAR,dighe2020latticebased,garg2021streaming}, the FTM task for touch-based invocation is extremely challenging because 1) the VA cannot rely on detection of a wake-word, and 2) directed speech is unconstrained and it could have very diverse acoustic and linguistic content~\cite{mallidi2018devicedirected,Gillespie_2020,norouzian2019exploring} -- see Table~\ref{tab:example}.

\begin{table}[]
\centering
\caption{Examples of device (un)directed utterances}
\begin{tabular}{|l|l|}
\hline
Device-directed   &Device-undirected  \\ \hline
Taxi cab service near me  &... I am not that hungry now ... \\ 
NBA playoffs  &... they said they would help ...  \\ \hline
\end{tabular}
\label{tab:example}
\end{table}

Training accurate FTM model requires a large number of training samples from device-directed and undirected classes. To obtain such data, human annotators must manually label large quantities of user audio, which is expensive, tedious and time-consuming. To this end, we introduce a simple yet effective approach that can be used to obtain training data in a weakly-supervised manner~\cite{zhou2018brief}, based on statistics from a small, anonymous, short-lived set of human labeled data. 
Specifically, we propose a deterministic approach (see Sec.~\ref{data}) based on signal-to-noise ratio (SNR) of the audio and an off-the-shelf text-based intent classification of the utterance to obtain weak labels on large amounts of unsupervised audio. This provides a large amount of training data that can then be used to train the target FTM models. Yet, since these data are weakly labeled, a model with enough learnable parameters can easily overfit the biases in the sampling process and is susceptible to the noise in the labels, resulting in poor generalization. 

As part of ongoing efforts to reduce the number of false positive activations by VAs, we tackle the challenges mentioned above using a combination of two FTM approaches: (i) ASR-based LatticeRNN FTM (LRNN)~\cite{Jeon_2019,ladhak2016latticernn}, a relatively small model ($\sim$5k tunable parameters), which uses low-dimensional ASR lattices and is less prone to overfitting (note that LRNN depends on a {\it pre-trained ASR} that has $\sim$60M parameters)~\cite{bengio2009learning}. (ii) Acoustics-only FTM (AFTM), a larger capacity self-attention based~\cite{vaswani2017attention,adya2020hybrid} model ($\sim$4.8M parameters), which consumes high-dimensional filter-bank features of the audio. We show that while this large model capacity can easily lead to overfitting on the weakly labeled training data, it also enables to successfully distill knowledge from LRNN during training, which is critical for improving its accuracy.

In the proposed approach, we first train a moderately accurate LRNN model using the weakly-labeled training data and use it to generate soft labels on the training data itself. Subsequently, we train a large AFTM model by supervised training using weak labels, and regularize its training using the notion of knowledge distillation~\cite{DBLP:journals/corr/HintonVD15} from the soft labels generated by LRNN. These soft labels provide AFTM model with instance level label smoothing \cite{YuanTLWF20,7780677} and we show that the distilled AFTM model obtains accuracy close to the teacher LRNN model. Moreover, the distilled AFTM significantly outperforms the baseline AFTM model trained using the weak labels but without the regularization via distillation. While the LRNN FTM depends on the computationally expensive full-fledged ASR, an accurate AFTM model can be deployed on devices with low-resource hardware where ASR can not be deployed. Finally, we show that a simple score-level fusion of the distilled AFTM and LRNN further improves the FTM accuracy, thereby showing the complementary nature of these two approaches: LRNN focuses on the high-level information such as sentence hypotheses for determining device-directedness of the audio, while AFTM focuses on low-level information such as presence of speech, background noise, and the general acoustic environment of the audio. The fusion approach is also promising for devices where both models can be deployed simultaneously. In what follows, we describe individual components of the proposed FTM approach, including the newly introduced approach to generate weakly-labeled training data, model training and distillation, and model fusion. Lastly, we outline the experimental setup and demonstrate the effectiveness of the proposed FTM in the target task. 

\section{Methods}
\subsection{Weakly-labeled Training Data}
\label{data}
Supervised training of accurate FTM models requires user utterances labeled as either intended or unintended towards the VA. However, labeling user-data for these categories requires manual annotation of large amount of data, which is slow, tedious and costly. While a small set of
data must be labeled to obtain a reliable development/test datasets, we refrain from labeling the training data for our FTM models. The key to our approach is to use weak-labeling of data without human-in-the-loop to obtain the training dataset~\cite{zhou2018brief}. 

\begin{figure}[htb]
\centering
\includegraphics[width=5cm, angle =0]{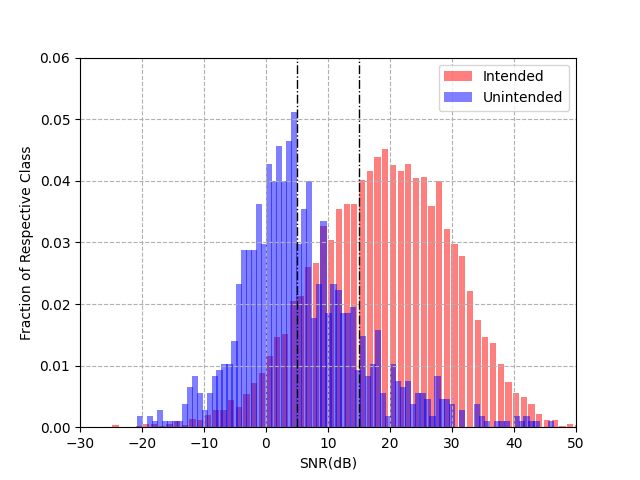}
\caption{{Distribution of the SNR signal on the development data.}}
\vspace{-5mm}
\label{fig:data}
\end{figure}

The weak-labeling is done using a deterministic approach (explained below) to grade user utterances as intended or unintended without graders having to listen to the audio or read the ASR transcript of the audio. Firstly, we note that typical intended queries are clean speech utterances which have a high signal-to-noise ratio (SNR), whereas false alarms often have noisy speech or simply background noise captured by the microphone. This is shown in Fig.~\ref{fig:data}, where the average SNR for intended utterances is $\sim$20dB, whereas for false alarms it is $\sim$0dB. The two categories have a large overlap in-between 5dB and 15dB SNR band. Secondly, we find that the intended queries are mostly questions or commands type of utterances, such as What time is it?, Set a timer, and so on. Therefore, the set of all intended queries loosely falls under a limited vocabulary and a question/command like grammatical sentence structure. A BERT \cite{devlin2018bert} style text-based intention classification model is used to weakly label utterances as background versus not background speech by processing the ASR output of the utterances. 

Based on the observations above, we design our deterministic approach for obtaining weak labels as follows:
\begin{enumerate}
    \item If a query has SNR$\leq$5dB and the text intent is classified as background speech, label it as unintended.
    \item If a query has SNR$\geq$15dB and the text intent is classified as not-background speech, label it as intended.
    \item If a query does not match any of the two conditions above, then we do not label or use such a query in our training data.
\end{enumerate}
We consider the labels obtained with the conditions defined above as weak/noisy labels. Note that this weak labeling strategy {\it cannot} be used to make decisions for {\it condition} 3 on test data, which covers $\sim$38\% of the data. Conversely, on the remaining data, we observe 8\% and 2\% misclassification rate when using the {\it conditions} 1 (unintended) and 2 (intended), respectively. This is expected due to the large overlap in the SNR distribution of the two classes (see Fig.~\ref{fig:data}). It also shows that the intent classification model is not perfect as it often acts on inaccurate ASR outputs in noisy speech conditions. On the other hand, the generated weakly-labeled data are well separated in terms of the intended/unintended class, yet, their labels are still noisy as they are not manually annotated. The lack of generalization and poor coverage on test data by this weak-labeling strategy motivates the training of the FTM models with the proposed regularization via distillation and model fusion, as effective ways of smoothing noise in the labels.

\begin{figure*}[t!]
\begin{minipage}[b]{.22\linewidth}
  \centering
  \centerline{\includegraphics[width=4.8cm]{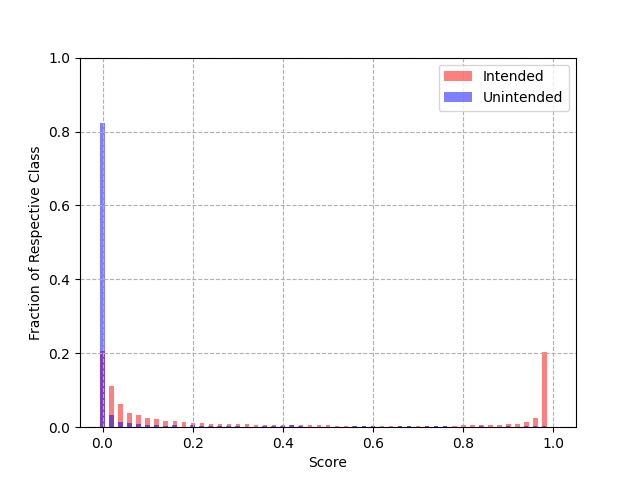}}
  \centerline{(a) AFTM}
\end{minipage}
\hfill
\begin{minipage}[b]{0.22\linewidth}
  \centering
  \centerline{\includegraphics[width=4.8cm]{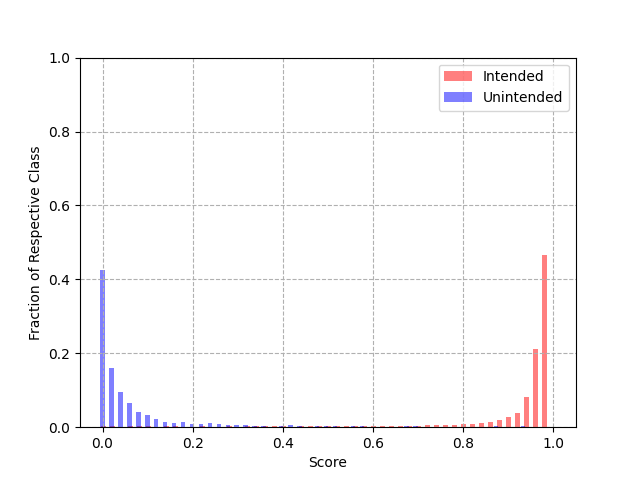}}
  \centerline{(b) LRNN}
\end{minipage}
\hfill
\begin{minipage}[b]{0.22\linewidth}
  \centering
  \centerline{\includegraphics[width=4.8cm]{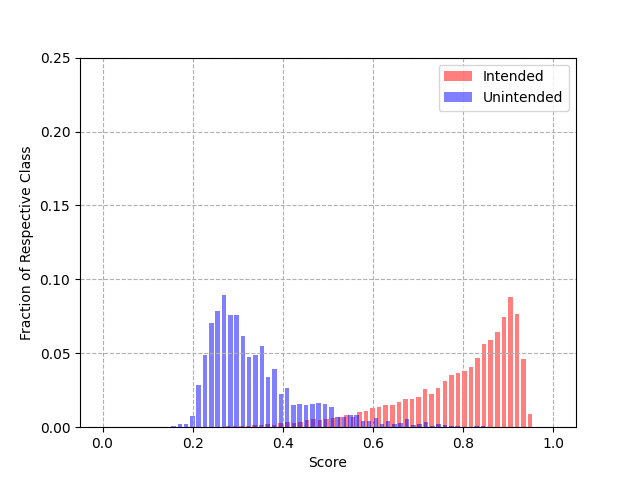}}
  \centerline{(c) AFTM-D}
\end{minipage}
\hfill
\begin{minipage}[b]{0.22\linewidth}
  \centering
  \centerline{\includegraphics[width=4.8cm]{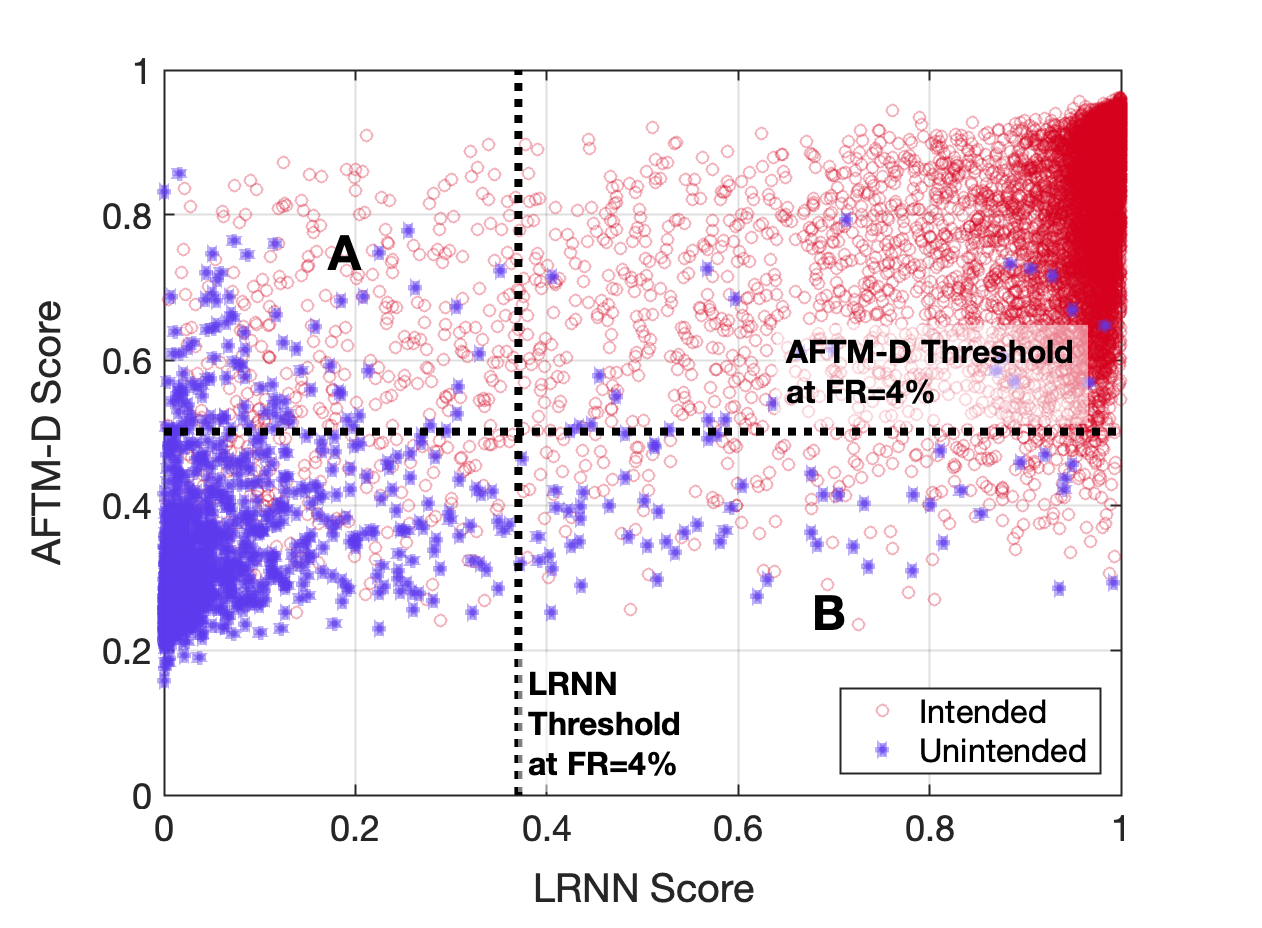}}
  \centerline{(d) LRNN and AFTM-D}
\end{minipage}
\caption{FTM score distribution for different models are depicted in (a)-(c). A scatter plot between LRNN scores and AFTM-D is shown in (d) which depicts their complementary nature in FTM task.}
\label{fig:scores}
\end{figure*}
\subsection{FTM Models}
\label{sec:ftm_models}

\subsubsection{Lattice RNN (LRNN)}
\label{lrnn}
The LRNN approach~\cite{Jeon_2019,ladhak2016latticernn} is based on ASR lattices obtained by decoding input audio using (i) posterior probabilities from an acoustic model (AM), and (ii) an external language model (LM), based on a weighted finite state transducer~\cite{mohri2002weighted}. The advantage of using LRNN in the FTM task is that device-directed speech is often less noisy and in-domain with its language model. Therefore, the \textit{best} sentence hypothesis has zero (or few) competing hypotheses in the decoded lattice graph. In contrast, false triggers are often noisy utterances that contain only background noise and/or speech, resulting in multiple competing ASR hypotheses in the lattice graph with no clear winner. The produced lattices are summarized using a RNN model as 16-D embeddings, which are then passed through a fully-connected layer (FCN) followed by Softmax to obtain the FTM scores for the target audio. While LRNN is effective in the FTM task, the full-fledged ASR needs to be run on device (which can be computationally expensive), or on servers (which increases the latency, which can affect the user experience). Also, the model needs be retrained each time the acoustic/language/ASR models change, since they affect the distribution of the lattices.
\subsubsection{Acoustic FTM (AFTM)}
\label{aftm}
In contrast to LRNN, the AFTM approach uses acoustic features only (see Sec.~\ref{exp}). The model is built upon a stack of self-attention (SA) layers~\cite{vaswani2017attention}, resembling the Encoder part of the base Transformer architecture~\cite{adya2020hybrid}. The SA layers effectively perform the task of feature extraction from input audio. Specifically, we compute 40-D mel-filterbank features from the audio at 100 frames-per-second (fps). At every time-step we splice 7 frames together to form a 280-D input window, followed by the sequence sub-sampling by a factor of 3. We use a stack of 6 self-attention layers (with 4 heads each), with each head yielding a 64-dimensional key, query and value vectors\footnote{The size of the hidden layer in the feed-forward (FF) layer is set to 1024, and we use skip connections for the SA/FF layers, followed by Layer Normalization in the output.}. Each head is concatenated to output a 256-D acoustic embedding, where every input produces a sequence of such embeddings. These are then summarized by a global attenion layer~\cite{luong-etal-2015-effective}, the output of which is passed through Softmax, to obtain the FTM score for the full audio. 

\subsection{Model Training and Inference}
\label{train_inf}
Both models are first trained individually by minimizing the Binary Cross Entropy (BCE) loss between the model predictions and the weak-labels (Sec.~\ref{data}). While the LRNN uses pre-defined acoustic and language models, the SA layers in the AFTM model are initialized using a pre-trained model on a large corpus of phonetic data, also used to train ASR. The phonetic classifier at the head of the acoustic model is replaced with the FTM classifier (i.e, the global attenion layer + softmax). Note that the phonetic data are not part of the FTM data corpus. We observe that the LRNN is more accurate than the base AFTM model in the target task (see Sec.~\ref{exp}). We attribute this to the fact that AFTM does not use a language model and ASR-decoding, which is critical when training on the weakly-labeled data. To this end, we introduce two model combination schemes that we find to be effective.

\subsubsection{Regularization via Distillation}
\label{models}
To use the robustness of the LRNN model, while keeping the simplicity of the AFTM, we use {\it knowledge distillation (KD)}~\cite{DBLP:journals/corr/HintonVD15,YuanTLWF20} to regularize the training of the AFTM model. Formally, the new loss function of the distilled AFTM (AFTM-D) is defined as:
\begin{equation}\label{kd_loss}
loss =  loss_{BCE} + \alpha*KL(LRNN,AFTM), 
\end{equation}
where the total {\it loss} of the AFTM-D is defined as the weighted sum of the BCE loss of the base AFTM model ($loss_{BCE}$), and the two-way instance-level KL divergence between the AFTM/LRNN probability scores on the training samples ($KL(f, g) = \sum_x f(x) log( g(x)/f(x))$). The LRNN scores are fixed in the regularization term, and the weight parameter $\alpha$ is tuned on the validation set ($\alpha=10$). We also found that this loss is more effective than using the pure KD approach (i.e., with no BCE loss). Furthermore, we also explored distilling the knowledge directly from the LRNN embeddings, similarly to~\cite{dighe2020knowledge} but on the full audio, using the mean-squared loss between the AFTM and LRNN embeddings, yet the simple KL-distillation from the LRNN scores was more effective in our experiments.  
\begin{figure}[htb]
\centering
\includegraphics[width=5cm]{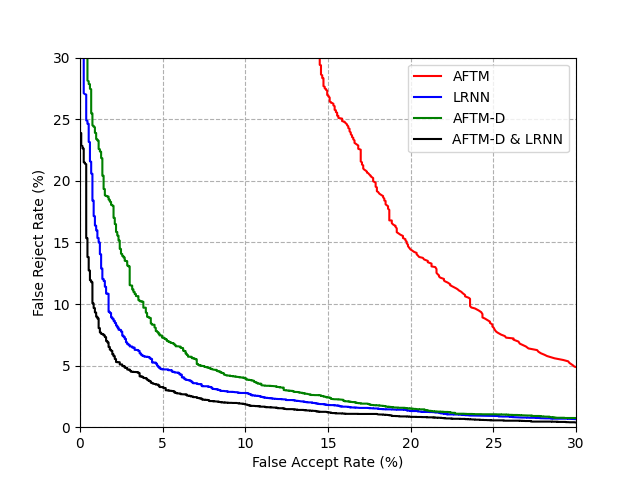}
\caption{DET curve showing accuracy of different models}
\label{fig:DET}
\end{figure}

\subsubsection{Model Fusion}
Since the AFTM and LRNN use different architectures for the underlying acoustic models, as well as a different inference mechanism (acoustic vs. lattice embeddings), we investigate the fusion of these two model scores to improve the robustness of the proposed FTM approach. This is motivated by the fact that ASR lattices focus on high-level information, such as word/sentence hypotheses, while the acoustic embeddings rely on low-level information, such as the presence of speech, background
noise, and the acoustic environment~\cite{dighe2020knowledge}. Instead of using the embedding-level fusion, as in~\cite{mallidi2018devicedirected}, we adopt a simple combination of the probability scores from the AFTM and LRNN. Specifically, we find that when the distilled AFTM model and LRNN are combined as:
\begin{equation}\label{fusion}
score_{fuse}=(score_{AFTM_D}+score_{LRNN})/2
\end{equation}
the FTM accuracy increases largely (see Sec.~\ref{exp}). This demonstrates the complementary information in the two models, even after the distillation. We believe this is for two reasons: (i) the AFTM-D and LRNN still focus on the low and high level information, respectively, due to their different inference mechanism. (ii) Since our training data are very noisy, the simple score fusion reduces the variance in the model predictions, which improves FTM accuracy on test data. Note that we also learned the weight parameters for the score fusion, but it did not further increase the accuracy, which supports our hypothesis in (ii).  

\section{Experiments}
\label{exp}
\subsection{Data and Evaluation Procedure}
We evaluate the FTM models on an anonymized and short-lived in-house dataset for touch-based invocation. Note that only the development/test sets are human-labeled, while the training data is weakly-labeled (as described in Sec.~\ref{data}). When the weak-labeling strategy is applied to the test partition, 18.7\% of data fall outside the SNR band we use for sampling, and on remaining data, it achieves 75\% accuracy in predicting the correct label. The available development set (used for obtaining the weakly-labeled data) is insufficient to train the FTM models, and it is needed to tune the model hyperparameters, as described below. For comparisons, we report the standard Detection Error Tradeoff (DET) curve, and Area Under the Curve (AUC).

\begin{table}[]
  \centering
  \caption{Accuracy of the FTM models}
        \begin{tabular}{|l|l|l|l|}
        \hline
         &{\bf EER[\%]} &{\bf FA@4\%FR} &{\bf AUC[\%]} \\ \hline
         {\it AFTM}           & 18.4 & 31.3 &  0.108\\ \hline
         {\it LRNN}           &  4.8 &  6.3 &  0.012\\ \hline\hline
         {\it AFTM-D}         &  6.2 &  9.9 &  0.016\\ \hline\hline
         {\it LRNN \& AFTM}   &  6.1 &  9.2 &  0.020\\ \hline
         {\it LRNN \& AFTM-D} &  4.0 &  3.8 &  0.007\\ \hline
        \end{tabular}
    \label{tab:DATA}
\end{table}
To assess the effects of the model regularization via distillation, we report the results for the base AFTM, LRNN and AFTM-D models. We further compare the fusion of these models (LRNN\&AFTM and LRNN\&AFTM-D). For all models, we used the Adam optimizer~\cite{kingma2014adam} with learning rates, gradient clipping norm, and dropout rate set on the development set. Note that all models perform FTM on full audios (by mapping the variable length acoustic signal to a fixed dimension embedding which is then transformed into a binary vector to compare against a predefined threshold (Sec~\ref{models}).

\subsection{Results}
Fig.~\ref{fig:scores} shows the distribution of the FTM scores for the intended/unintended class for different models. All models are trained on the weakly-labeled data (Sec.~\ref{data}). The baseline AFTM model (Fig.~\ref{fig:scores}(a)), which consumes acoustic features directly, is unable to generalize well on the test set as it tends to misclassify a large number of examples from the intended class. On the other hand, LRNN (Fig.~\ref{fig:scores}(b)), which relies on ASR, separates well the two classes. However, the proposed strategy of regularization via distillation results in the AFTM-D model (Fig.~\ref{fig:scores}(c)), the scores of which produce the FTM accuracy close to that of LRNN (see below). Finally, in Fig.~\ref{fig:scores}(d), we show the scatter plot of the scores by these two models. Note that, for instance, the unintended examples (in {\it blue}) falsely classified by AFTM-D, and correctly classified by LRNN in the region denoted as $A$. Conversely, the unintended examples in the region denoted as $B$ are falsely classified by LRNN, and correctly classified by AFTM-D. Similar observations can be made for the misclassified examples from the intended class, where the models exhibit the opposite behaviour in these two regions. This complimentary nature allows us to improve upon the models' individual accuracy using a simple 1-D classifier (by averaging the models’ scores), as proposed here. 

Fig.~\ref{fig:DET} depicts the DET curve on our test set, with false-reject-rate (FRR) and false-accept-rate (FAR) set on the y/x axes, respectively. We immediately notice the large improvement in the AFTM accuracy (due to the introduced model regularization via distillation from LRNN) across all operating regions. This confirms our hypothesis that the pure acoustic model is unable to efficiently deal with weakly-labeled data due to the high noise in their labels. On the other hand, the LRNN achieves high accuracy. This is in part due to its low-capacity (only $\sim$5k tunable parameters) and low dimensional input features, making it more difficult to overfit noisy labels~\cite{bengio2009learning}. It also uses a LM and ASR decoding to reduce the hypothesis space, which helps further in dealing with noisy labels. The AFTM model capacity allows distillation of this knowledge during training, bringing its accuracy closer to LRNN, with no need to use ASR during inference time. 

By looking at the proposed fusion of the LRNN and AFTM models, we note that the combination with the base AFTM only hurts the FTM accuracy (as the majority of its predictions are inaccurate). On the other hand, the fusion between LRNN and AFTM-D further improves the overall FTM accuracy. Based on this, we conclude that the base AFTM does not bring extra information to the already highly accurate LRNN model. However, since LRNN produces more skewed scores (see Fig.~\ref{fig:scores}), the fusion with AFTM-D ``softens'' these overly confident scores by adding additional emphasis on the acoustic evidence. We also attribute the improvements in the FTM accuracy to the fact that  combining these two complementary models reduces the noise variance in the final prediction, given that our training data is highly noisy.

Table~\ref{tab:DATA} shows EER, FAR at a hypothetical FRR of 4\%, and normalized AUC for the FTM models. Note the effects of the proposed regularization of the base AFTM model via distillation from LRNN: the accuracy of AFTM-D improves $\sim$66\% over the base AFTM, across all three metrics. Compared to LRNN, AFTM-D achieves $\sim$30\% lower accuracy, however, it can easily be run on low-resource hardware. On the hardware that permits deploying both models (LRNN\&AFTM-D), we see an improvements of $\sim$20\% in EER, and $\sim$40\% in FAR/AUC, over the (single) LRNN model. 

\section{Conclusions}
We have addressed the challenging problem of detection of device-directed queries from the keyword free utterances. To this end, we introduced a novel data sampling strategy for training detection models, which do not require manual annotation of the target audios, thus, largely reducing the data annotation effort. To achieve effective mitigation of device-undirected queries, we proposed two modeling approaches to this task (regularization via distillation and model fusion). We showed in our experiments that applying these is critical when dealing with weakly-labeled data, through combination of acoustic and ASR-based models. The proposed approach limits largely the need for data labeling, which is important for large-scale training of FTM models.

\section{Acknowledgements}
We thank S. Vishnubhotla and B. Theobald for their feedback and support during the course of this work.

\bibliographystyle{IEEEtran}

\bibliography{mybib}

\end{document}